\documentclass[11pt]{article}
\usepackage{amssymb}
\usepackage{graphics}
\usepackage{epsfig}
\usepackage{a4wide}
\usepackage{slashed,cite}
\usepackage{color}

\begin{document}

\newcommand{\nn}{\nonumber}
\newcommand{\eewwr}{$e^+e^- \to W^+W^-\gamma~$}
\newcommand{\decayeewwr}{$e^+e^- \to W^+W^-\gamma \to l_1^+ l_2^- \nu_{l_1}\bar {\nu}_{l_2}\gamma~$}
\newcommand{\wwr}{$W^+W^-\gamma~$}

\title{Electroweak radiative corrections to $W^+W^-\gamma~$ production at the ILC}
\author{ Chen Chong, Ma Wen-Gan, Zhang Ren-You, Zhang Yu, Chen Liang-Wen, and Guo Lei   \\
{\small  Department of Modern Physics, University of Science and Technology of China (USTC),}  \\
{\small   Hefei, Anhui 230026, P.R.China}}

\date{}
\maketitle \vskip 15mm
\begin{abstract}
We provide and discuss the precision predictions for the $W^+W^-\gamma~$ production at the ILC including the full electroweak (EW) one-loop corrections and high order initial state radiation (ISR) contributions in the Standard Model. The dependence of the leading order (LO) and EW corrected cross sections on the colliding energy is investigated. We find that the EW correction suppresses the LO cross section significantly, and the ISR effect beyond ${\cal O}(\alpha)$ is important near the threshold, but is negligible in the high energy region. We provide the LO and EW corrected distributions of the transverse momenta and rapidities of final $W^-$-boson and photon as well as the $W$-pair invariant mass. From the various kinematic distributions, we find that EW correction strongly depends on the final state phase space. We investigate the leptonic decays of the final W-boson pair by adopting the narrow width approximation (NWA), and find that the final produced photon and leptons can be well separated from each other.
\end{abstract}


\vfill \eject \baselineskip=0.32in

\renewcommand{\theequation}{\arabic{section}.\arabic{equation}}
\renewcommand{\thesection}{\Roman{section}.}
\newcommand{\nb}{\nonumber}

%slash:
\newcommand{\Dir}{\kern -6.4pt\Big{/}}%su lettere italiane minuscole
\newcommand{\Dirin}{\kern -10.4pt\Big{/}\kern 4.4pt}
\newcommand{\DDir}{\kern -7.6pt\Big{/}}%su lettere italiane maiuscole
\newcommand{\DGir}{\kern -6.0pt\Big{/}}%su lettere greche

\makeatletter      % '@' is now a normal "letter" for TeX
\@addtoreset{equation}{section}
\makeatother       % '@' is restored as a "non-letter" character for TeX

\par
\section{Introduction}
\par
The Standard Model (SM) has achieved great success in providing a remarkably accurate descriptions of the existing high energy data. Particularly, a tremendous achievement is the discovery of a $126~{\rm GeV}$ long-sought Higgs boson predicted by the SM which was announced in July 2012 by both the ATLAS and CMS collaborations at the LHC \cite{Higgs}. The main goals of the forthcoming experiments are to understand the nature of the Higgs boson and discover the signature of new physics beyond the SM.

\par
Theoretically the SM gauge invariance provides stringent constraints on the strengths of gauge couplings, which reveal the gauge structure in the SM. Moreover, because the longitudinal components of massive gauge bosons ($W^{\pm}/Z$) originate from the spontaneous symmetry breaking, accurately testing the gauge couplings is useful to either confirm the electroweak (EW) symmetry breaking mechanism or indicate new physics beyond the SM.

\par
The multiple gauge boson productions are especially important in probing the self-coupling properties of the gauge bosons, and would give a crucial test of the non-Abelian structure of the SM. For the direct study of the quartic gauge couplings (QGCs) of EW bosons, the measurements of the triple gauge boson productions are required. In the last few years, the calculations of triple gauge boson productions at hardron colliders up to the QCD next-to-leading order (NLO) have been completed \cite{zzzqcd, wwzqcd, www-zzwqcd,  tribosonqcd, wwr-zzrqcd, wzrqcd, wrrqcd-U.B, wrrqcd-G.B, zrr-rrrqcd}. Recently, the NLO EW correction to the $WWZ$ production at the LHC was also presented \cite{wwzqcdew}. The experimental studies for $WW\gamma$ and $WZ\gamma$ productions and constraints on anomalous QGCs at the $\sqrt{s} = 8~{\rm TeV}$ LHC are provided in Ref.\cite{CMS-wwr}. At present there is no evidence for existing anomalous QGCs from current data. Due to the heavy background at hadron collider, the triple gauge boson productions at the future International Linear Collider (ILC) are much cleaner than at hardron machines \cite{ILC}. Therefore, the precision theoretical understand of these processes at the ILC at least to one-loop order is necessary. Up to now, the NLO EW corrections to $WWZ$, $ZZZ$ and $Z\gamma\gamma$ productions at the ILC were provided in Refs.\cite{zzzew,wwzew,zzz-wwzew,zrr-ew}.

\par
Since the $W^+W^-\gamma$ production at the ILC can be used to explore the $W^+W^-\gamma\gamma$ and $W^+W^-Z\gamma$ QGCs, the precision understanding of this process is important. The effects of anomalous QGCs in $W^+W^-\gamma$ production at the LEP, ILC and CLIC were theoretically investigated in Refs.\cite{AQGCS1,AQGCS2}. The phenomenological study on the \eewwr process at the leading order (LO) in the SM was presented in Refs.\cite{ee->wwr1,ee->wwr2,ee->wwr3}, while the NLO EW correction to \eewwr is still missing which would be indispensable to match the ILC experimental accuracy.

\par
In this paper, we present the full NLO EW corrections to the \eewwr process in the SM at the ILC, as well as the high order initial state radiation (ISR) contributions at the leading-logarithmic approximation in the structure function method. The leptonic decays of the unstable $W$-boson pair is also investigated by adopting the narrow width approximation (NWA).  The rest of the paper is organized as follows: In section 2, we give the details of the calculations of the LO and EW corrections to \eewwr process. In section 3, numerical results and discussion are given. Finally we make a short summary.

\vskip 5mm
\section{Calculation setup}
\par
\begin{figure}
\begin{center}
\includegraphics [ scale = 1.0]{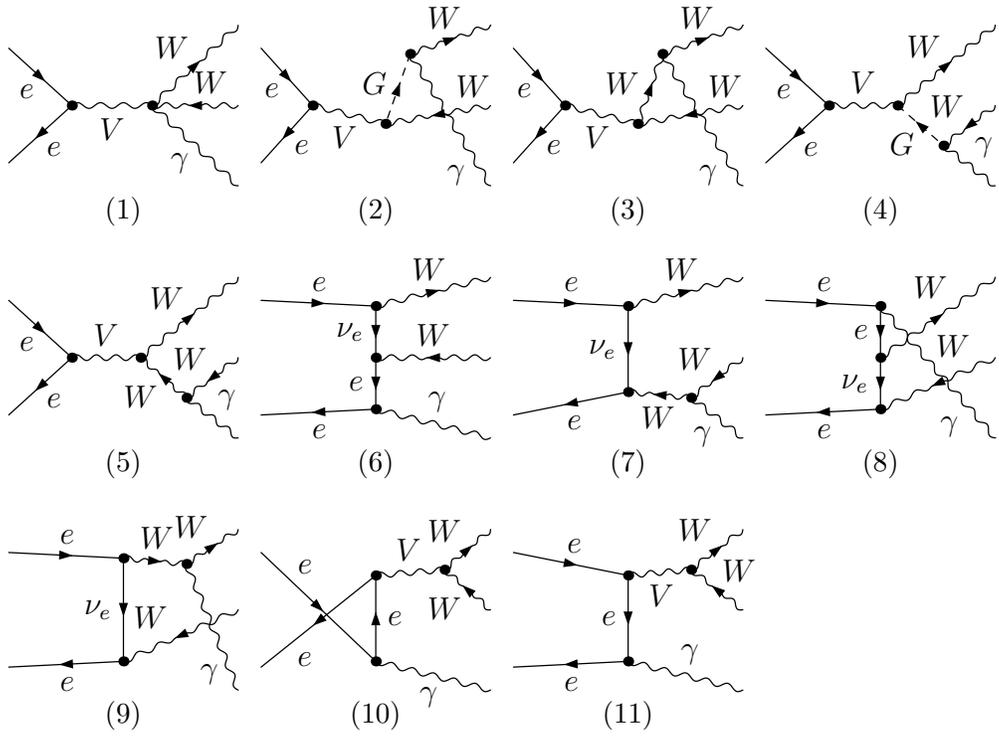}
\caption{\label{fig-lo} The LO Feynman diagrams for the \eewwr process, where V represents $Z$ or $\gamma$. }
\end{center}
\end{figure}

In our calculation, we apply FeynArts-3.9 \cite{FeynArts} to generate automatically the Feynman diagrams. The amplitudes are given in the 't Hooft-Feynman gauge and are subsequently reduced by using FormCalc-7.4 \cite{FormCalc,FC-LT}. Due to the smallness of the electron mass, we ignore the contributions from the graphs involving Higgs/Goldstone-electron-positron Yukawa interactions. The Feynman diagrams for the \eewwr process are depicted in Fig.\ref{fig-lo}.
%%%%%%%%%fig-lo%%%%%%%%%%%%%%%%%%

\par
The LO cross section and the NLO EW correction for \eewwr are of the ${\cal O}(\alpha^3)$ and ${\cal O}(\alpha^4)$, respectively. Traditionally, the fine structure constant $\alpha$ is defined from a full $e$-$e$-$\gamma$ coupling for on-shell external particles in the Thomson limit, leading to the renormalized value $\alpha=\alpha(0)$ in $\alpha(0)$-scheme. This definition is not the most appropriate since many processes take place at the weak scale or higher and the corrections due to the running of $\alpha$ are sensitive to light fermion masses $m_f$ through logarithms ${\rm log}(Q^2/m_f^2)$ with a typical scale $Q$ of the process. For processes with a LO cross section of ${\cal O}(\alpha^n)$ and containing $l$ external photons, the logarithms resulting from the charge renormalization can be canceled by the ones in the corresponding external photon wave-function counterterm. When $l$ equals $n$, the cancel is complete and the EW correction is free of the logarithms. If $l$ is less than $n$, the logarithm terms remain, but can be absorbed into the running $\alpha$ by using $\alpha(M_Z^2)$-scheme or $G_F$-scheme. This statement expresses the fact that the proper coupling of a real photon should be $\alpha(0)$ \cite{Les-Houches}. In our calculation, we use a mixed input-parameter scheme, i.e., the couplings related to the external photons are fixed with $\alpha=\alpha(0)$ and the others with $\alpha=\alpha_{G_F}= \frac{\sqrt2 G_F M_W^2}{\pi}(1-\frac{M_W^2}{M_Z^2})$. Accordingly the LO cross section is of ${\cal O}(\alpha_{G_F}^2\alpha(0))$ and the NLO EW corrections are of ${\cal O}(\alpha_{G_F}^2\alpha(0)^2)$.

\par
\subsection{Virtual corrections}
\par
The NLO EW virtual corrections to the \eewwr include 2485 diagrams, and we divide them into self-energy (801), triangle (983), box (477), pentagon (108) and counterterm (116) graph groups. In the one-loop diagrams, there are both ultraviolet (UV) and infrared (IR) divergences. The definitions of the relevant renormalization constants using the on-mass-shell conditions are presented in Ref.\cite{EW-RC}. The UV divergences are regularized using the dimensional regularization scheme and can be canceled exactly after performing the renormalization procedure. Because some universal corrections have been absorbed in $\alpha_{G_F}$, we have to subtract this part from the virtual corrections calculated in the $\alpha(0)$-scheme to avoid double counting. The electric charge renormalization constant in the $G_F$-scheme is modified as $\delta Z_e^{G_F}=\delta Z_e-\frac{1}{2}\Delta r$, where $\Delta r$ is given by considering the one-loop EW corrections to the muon decay \cite{deltar}. In our calculation, the corresponding $\Delta r$ term should be subtracted twice because there are two $\alpha_{G_F}$ couplings in the LO cross section. We also adopt the dimensional scheme to regularize the IR singularities. After adding the contribution of real photon emission process, the soft IR divergences in loops are canceled and the final result is IR finite.
%%%%%%%%%fig-loop%%%%%%%%%%%%%%%%%%
\begin{figure}
\begin{center}
\includegraphics [ scale = 1.0]{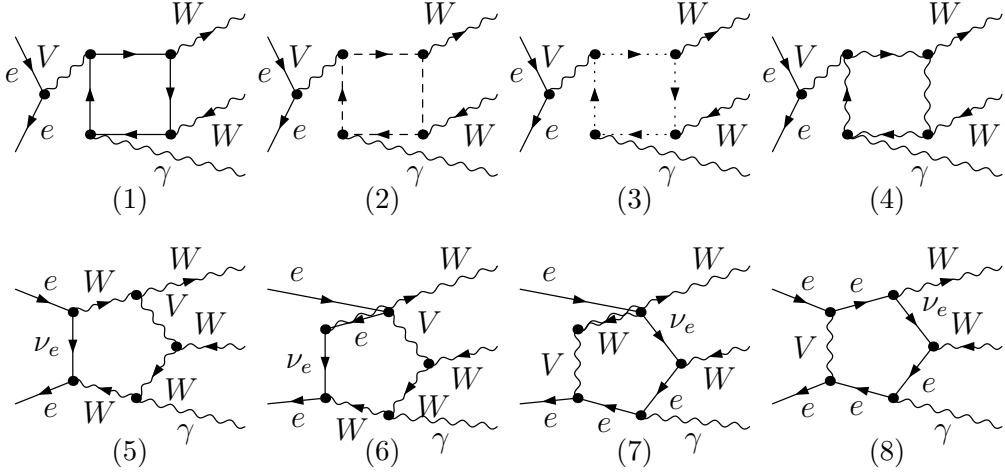}
\caption{\label{fig-loop} Some representative box and pentagon diagrams of rank 4 for the \eewwr process, where $V=Z/\gamma$. The solid, dash, dot and wave lines stand for fermion, scalar, ghost and vector boson, respectively.}
\end{center}
\end{figure}

\par
In the calculation of loop Feynman amplitudes, the $n$-point tensor integrals ($n \leq 4$) are reduced to scalar integrals recursively by using the Passarino-Veltman algorithm \cite{PV} and the 5-point integrals are decomposed into 4-point integrals by using the method of Denner and Dittmaier \cite{Denner-5-4}. In our previous work \cite{ZZW-LED}, we addressed the numerical instability resulting from the small Gram determinant ($detG$), which can be solved by using the method analogous to that in Refs.\cite{wwzqcdew,zzz-wwzew}. Except the above instability, we encounter another numerical problem originating from the scalar one-loop four point integrals which is  described detailedly in Ref.\cite{zzz-wwzew}. In our calculation the scalar integrals are evaluated by using the library LoopTools-2.8 \cite{FC-LT}, which provides two versions, i.e., one based on the FF package \cite{FF} and another one based on the program implementated by Denner \cite{Denner-scalar}. Any one of them alone is not good enough for our calculation and makes the numerical problem to be much more serious. Therefore, we choose the repaired second version as our default version, except some special cases where it fails and the first version is used. To check the correctness of our codes, we compare the scalar integral results calculated using the modified LoopTools-2.8 with those using OneLoop \cite{oneloop}, and find they are coincident with each other. In Fig.\ref{fig-loop}, we depict some representative box and pentagon diagrams of rank 4, which are the most complicated diagrams in topology involved in our calculation.

\par
\subsection{Real photon emission corrections}
\par
The real emission process contains an extra photon radiating off the initial or final states. We introduce two cutoff parameters $\delta_s$ and $\delta_c$ by adopting the two cutoff phase space slicing (TCPSS) method \cite{tcpss-isr}, and divide the phase space of the photon bremsstrahlung process into soft ($E_{\gamma} \leq \delta_s \sqrt{s}/2$), hard collinear  ($E_{\gamma} > \delta_s \sqrt{s}/2$ and $\cos\theta_{\gamma e} \leq 1 - \delta_c$) and hard noncollinear ($E_{\gamma} > \delta_s \sqrt{s}/2$ and $\cos\theta_{\gamma e} < 1 - \delta_c$) regions, where $\theta_{\gamma e}$ is the angle between the  bremsstrahlung photon and the electron/positron. Then the cross section for the real emission process $e^+ e^- \to W^+ W^- \gamma \gamma$ is decomposed as
\begin{eqnarray}\label{eq-1}
d \sigma_{{\rm real}} = d \sigma_{{\rm soft}}(\delta_s) + d \sigma_{{\rm coll}}(\delta_s, \delta_c) + d \sigma_{{\rm noncoll}}(\delta_s, \delta_c).
\end{eqnarray}
The soft correction $d \sigma_{{\rm soft}}$ contains soft IR singularities that can be canceled exactly by those in the virtual corrections. Since the initial electron, positron and the final $W^{\pm}$-bosons are all fundamental particles with non-zero mass, there is no collinear IR singularity in the hard collinear region. However, the smallness of the electron mass induces the quasi-collinear IR divergences from the photon radiation off the incoming electron/positron, i.e., initial state radiation (ISR). The hard noncollinear correction $d \sigma_{{\rm noncoll}}$ is IR finite, and therefore the phase space integration can be performed numerically by using Monte Carlo technique. All the three parts of the real photon emission correction in Eq.(\ref{eq-1}) separately depend on the cutoff parameters $\delta_s$ and $\delta_c$, but the total real photon emission correction $d \sigma_{{\rm real}}$ should be independent on the cutoff parameters. In our numerical calculation, we take $\delta_c = \delta_s/50$ and verify the cutoff independence of the real photon emission correction in the range of $\delta_s \in [10^{-4}, 10^{-2}]$.

\par
\par
We also use the dipole subtraction (DS) method \cite{dipole-qed} to extract the IR singularities of the real photon emission correction for comparison. In this method dipole terms are introduced to approximate the squared amplitude in the soft/collinear region for the real photon emission process. That means
$d \sigma_{{\rm real}}-d \sigma_{{\rm dipole}}$ is IR finite and can be integrated numerically. In order to leave the total result unchanged, the subtracted terms have to be added back after analytical integration of the real emission particle phase space. The formalism of the subtracted dipole terms can be built up from a process independent approach, which was first presented for QCD with massless unpolarized partons by Catani and Seymour \cite{dipole-qcd} and was subsequently generalized to photon radiation from massive or massless fermions by Dittmaier\cite{dipole-qed}. Photon radiation from charged bosons has the same IR singular structure as fermions, therefore we use the general subtraction formalism presented in Ref.\cite{dipole-qed} directly. We also check the independence on the parameter $\alpha$, which is introduced to control the size of dipole phase space \cite{dipole-a}. Technically, we use TCPSS method in our following calculation taking its advantage of clear physics picture, and the DS method is used to verify the correctness of our numerical calculation.

\par
The ISR quasi-collinear IR divergences can be canceled partially by those in the virtual contributions. The left quasi-collinear divergences would lead to large radiative corrections of the form $\alpha^n{\rm log}^n(m_e^2/Q^2)$ at the leading logarithmic (LL) level. To achieve the precision at the $0.1\%$ level, the contributions of this part beyond ${\cal O}(\alpha)$ have to be taken into account. By using the structure function method \cite{isr,tcpss-isr}, the ISR effect is written as the convolution of the LO cross section with structure functions:
\begin{eqnarray} \label{ISR1}
\int d\sigma_{\rm ISR-LL}~=~\int_0^1 dx_1 \int_0^1 dx_2~\Gamma_{ee}^{\rm LL}(x_1,Q^2)
~ \Gamma_{ee}^{\rm LL}(x_2,Q^2) \int d\sigma (x_1p_{e^-},x_2p_{e^+}),
\end{eqnarray}
where $x_{1,2}$ denote the fractions of the momentum carried by the electron and positron after photon radiation, $Q$ is the typical scale of the hard scattering process chosen as $\sqrt{s}$ in our calculation, and $\Gamma_{ee}^{\rm LL} (x,Q^2)$ is the LL structure function. The contributions to $\Gamma_{ee}^{\rm LL} (x,Q^2)$ originate from two parts: the soft photon part which can be resumed to an exponential term, and the hard photon part which has to be calculated order by order. In Ref.\cite{isr}, the explicit expression for $\Gamma_{ee}^{\rm LL} (x,Q^2)$ up to the ${\cal O}(\alpha ^3)$ is given. Note that the ISR effect at the ${\cal O}(\alpha)$ has been contained in the radiative corrections described before. When adding Eq.(\ref{ISR1}) to the NLO EW corrected result, we have to subtract the lowest-order and one-loop contributions to avoid double counting. The formulas for the subtracted terms are presented in Ref.\cite{tcpss-isr}. In the following, the subtracted ISR effect is called as high order ISR (h.o.ISR) contribution beyond ${\cal O}(\alpha)$ and the summation of all the parts mentioned above as EW corrected result.

\vskip 5mm
\section{Numerical results}
\par
In this section, we present and discuss the numerical results for the LO and EW corrected observables for the \eewwr process. The SM parameters used in our calculation are taken as \cite{pdg}:
\begin{equation}
\arraycolsep 2pt
\begin{array}[b]{lcllcllcl}
 G_F~=~1.1663787 \times 10^{-5},~~\alpha (0)~=~1/137.035999074,  \\
 M_{W}~= ~80.385~{\rm GeV}, ~~M_{Z}~= ~91.1876~{\rm GeV},~~M_H~=~126~{\rm GeV}, \\
 m_e~=~0.510998928~{\rm MeV},~~m_{\mu}=105.6583715~{\rm MeV},~~m_{\tau}~=~1.77682~{\rm GeV}, \\
 m_u~=~66~{\rm MeV},~~m_c~=~1.2~{\rm GeV},~~m_t~=~173.07~{\rm GeV}, \\
 m_d~=~66~{\rm MeV},~~m_s~=~150~{\rm MeV},~~m_b~=~4.3~{\rm GeV},
\end{array}
\label{SMpar}
\end{equation}
where the light quark (all quarks except $t$-quark) masses are adjusted to reproduce the hadronic contribution to the photonic vacuum polarization \cite{q-mass}. The Cabibbo-Kobayashi-Maskawa matrix is set to be unit matrix. As discussed in Sect.II, we use a mixed scheme to get the value of the fine structure constant. The LO cross section is of ${\cal O}(\alpha_{G_F}^2\alpha(0))$ and the NLO EW corrections are proportional to $\alpha_{G_F}^2\alpha^2(0)$.

\par
The final state for the $W^+ W^- \gamma$ production contains only one photon at the LO, while contains at most two photons up to the EW NLO. For a two-photon event originating from the hard noncollinear region of the real emission process, we apply the Cambridge/Aachen (C/A) jet algorithm \cite{jet-CA} to photon candidates. That means when the resolution of the two final photons satisfies the constraint of $R=\sqrt {\Delta y^2+\Delta \phi^2}<0.4$, where $\Delta y$ and $\Delta \phi$ are the differences of rapidity and azimuthal angle between the two photons, we merge them into one new photon with momentum $p_{ij,\mu}=p_{i,\mu}+p_{j,\mu}$ and call it as a ``one-photon'' event, otherwise it is called a ``two-photon'' event. In our calculation, we collect the ``one-photon'' events with the constraints of $p_T^{\gamma}>15{\rm GeV}$ and $\vert y_{\gamma}\vert\le 2.5$ to exclude the IR divergence in the LO calculation. For the ``two-photon'' events, at least one photon is required to satisfy the former constraints for ``one-photon'' event. When both two photons pass the transverse momentum and rapidity cuts, we label them as the leading photon and next-to-leading photon according to the criterion of $p_T^{\gamma,L}>p_T^{\gamma,NL}$.

\par
\subsection{Total cross sections}
\par
\begin{figure}%[htbp]
\begin{center}
\includegraphics[scale=0.3]{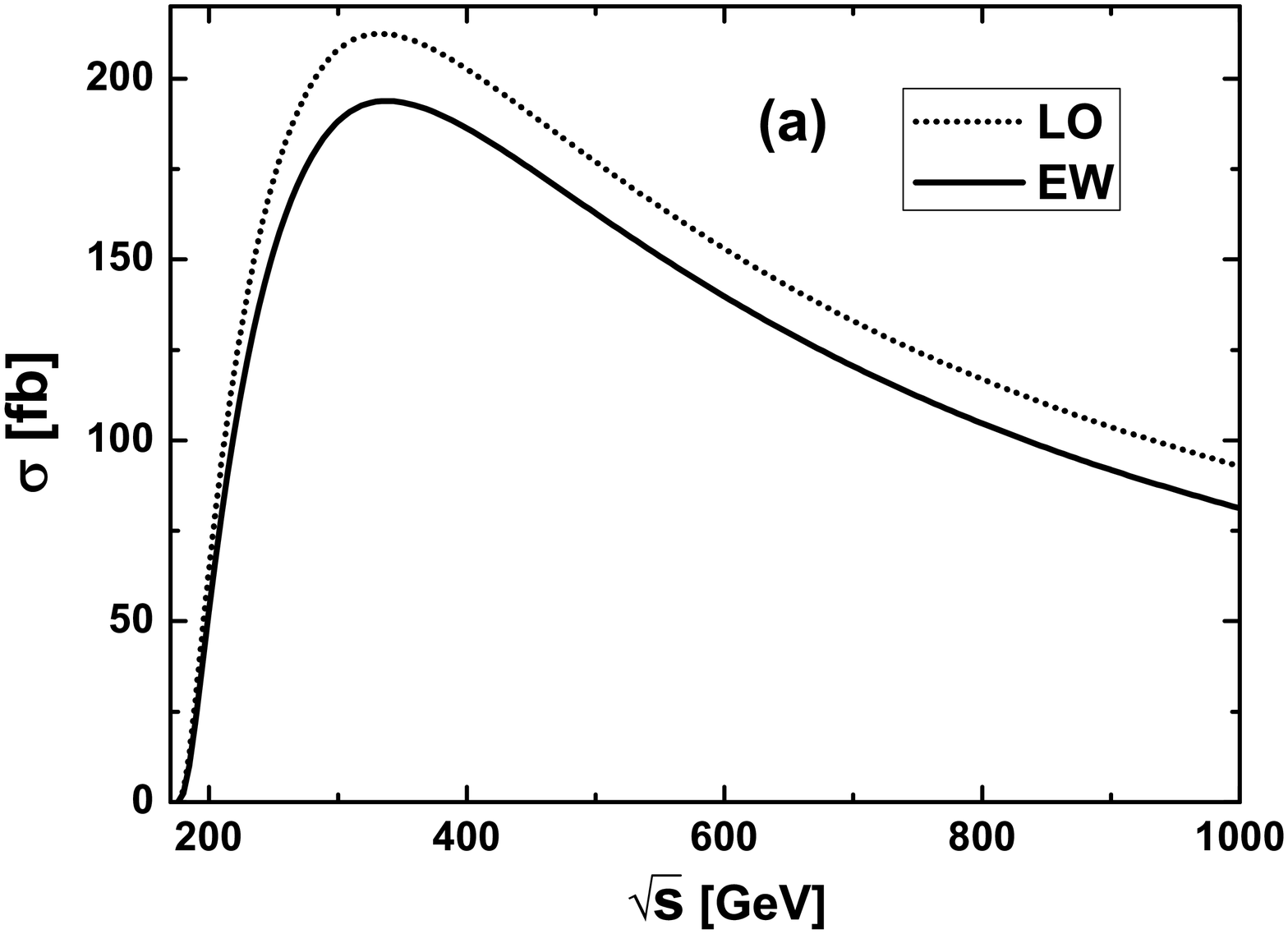}%
\includegraphics[scale=0.3]{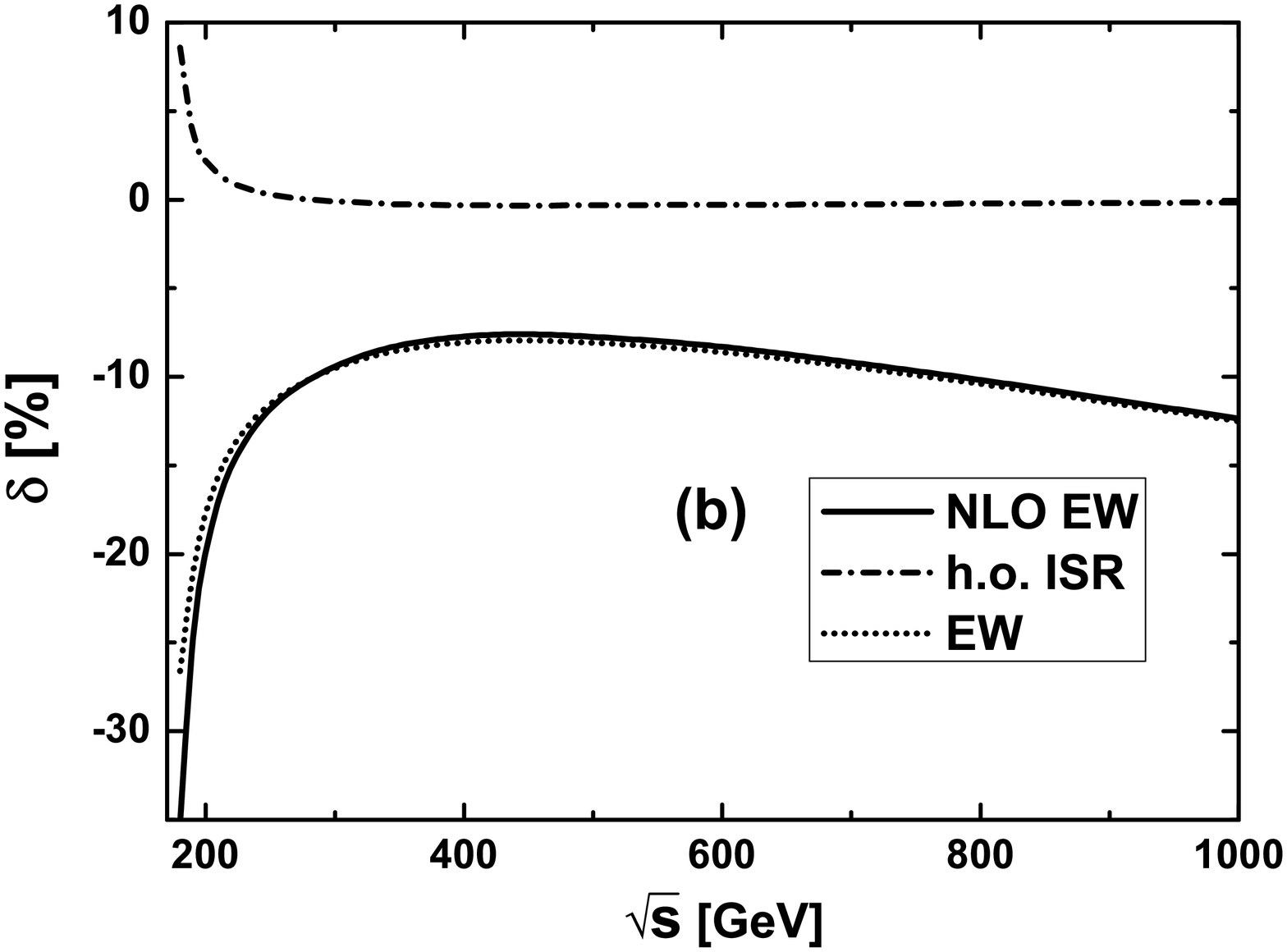}%
\caption{\label{sqrts} (a) The LO and EW corrected cross sections for the \eewwr process in the SM. (b) The corresponding NLO EW, h.o. ISR and EW relative corrections. }
\end{center}
\end{figure}

In Fig.\ref{sqrts}(a) we present the dependence of the LO and EW corrected integrated cross sections on the colliding energy $\sqrt{s}$ for the \eewwr process in the SM. The corresponding NLO EW, h.o. ISR and EW relative corrections, defined as $\delta \equiv \frac{\sigma-\sigma_{LO}}{\sigma_{LO}}$, are shown in Fig.\ref{sqrts}(b). From the figures, we find that both the LO and EW corrected integrated cross sections are sensitive to the colliding energy, and reach their maxima at the position of $\sqrt s\sim 330~{\rm GeV}$. The EW correction suppresses the LO cross sections in the whole plotted $\sqrt s$ region. As indicated in Fig.\ref{sqrts}(b), the absolute NLO EW relative correction becomes very large near the threshold. This is due to the Coulomb singularity effect coming from the instantaneous photon exchange in Feynman loops which has a small spatial momentum. At high energies the absolute NLO EW relative correction is also significant and goes up slowly with the increment of colliding energy. This behaviour is typical because we don't include the weak boson emission process in our calculation and the Sudakov logarithms like $\alpha {\rm ln}^2(s/M_W^2)$ dominate the weak corrections at high energies. Fig.\ref{sqrts}(b) shows that the ISR effect beyond ${\cal O}(\alpha)$  is distinct near the threshold, (e.g., the relative correction is $8.58\%$ at $\sqrt s = 180~{\rm GeV}$), but decreases to be lower than $0.5\%$ in the region of $\sqrt{s} > 250~{\rm GeV}$ which can be negligible. To show the results more explicitly, we present some representative numerical results of the LO, EW corrected cross sections, and the corresponding NLO EW, h.o. ISR and EW relative corrections in Table.\ref{tab-sqrts}.
%%%%%%%%%%%%%%fig-3%%%%%%%%%%%%%%%%%%%%%%%%%%%%%
%%%%%%%%%%%%%%fig-3%%%%%%%%%%%%%%%%%%%%%%%%%%%%%
\begin{table}
\center
\begin{tabular}{cccccc}
\hline \hline
$\sqrt{s}(GeV)$ & $\sigma_{LO}(fb)$
&$\sigma_{EW}(fb)$ & $\delta_{NLO}(\%)$& $\delta_{h.o.ISR}(\%)$& $\delta_{EW}(\%)$  \\
\hline \hline
180   & 3.319(1)    & 2.436(1)     &-35.20      & 8.58     & -26.62 \\
190   & 30.904(3)   & 24.408(8)    &-24.71      & 3.70     & -21.01 \\
200   & 62.316(7)   & 52.928(15)   &-19.91      & 2.20     &-17.71  \\
250   & 172.23(3)   & 152.41(6)    &-11.79      & 0.28     & -11.51 \\
300   & 208.00(3)   & 188.23(8)    &-9.406      & -0.099   & -9.505 \\
350   & 211.43(3)   & 193.47(9)    &-8.23       & -0.26    & -8.49  \\
400   & 202.62(3)   & 186.31(9)    &-7.73       & -0.32    & -8.05  \\
500   & 177.02(3)   & 162.74(9)    &-7.75       & -0.32    & -8.07  \\
800   & 116.87(2)   & 104.72(9)    &-10.18      & -0.22    &-10.40  \\
1000  & 92.87(2)    & 81.26(9)     &-12.35      & -0.15    & -12.50 \\
\hline  \hline
\end{tabular}
\caption{ \label{tab-sqrts} The LO, EW corrected cross sections ($\sigma_{LO}$, $\sigma_{EW}$), and the corresponding NLO EW, h.o. ISR and EW relative corrections ($\delta_{NLO}$, $\delta_{h.o.ISR}$ and $\delta_{EW}$) for the \eewwr process in the SM.}
\end{table}
%%%%%%%%%%%%%%%%%%%%

\par
\subsection{Kinematic distributions}
\par
In this section we investigate the EW corrections to the kinematic distributions, where the colliding energy is taken as $\sqrt{s}=500~{\rm GeV}$ and the differential relative EW correction is defined as $\delta(x) \equiv \left(\frac{d\sigma_{EW}}{dx}-\frac{d\sigma_{LO}}{dx}\right)/\frac{d\sigma_{LO}}{dx}$. Due to the CP conservation in the SM, the transverse momentum distribution of $W^+$ is the same as $W^-$ and the rapidity distribution of $W^+$ at $e^+e^-$ colliders can be built up from the one of $W^-$ by reversing the curve from left to right. In the following plots we only provide the distributions of $p_T^{W^{-}, \gamma}$, $y^{W^{-},\gamma}$ and $W$-pair invariant mass $M_{WW}$.

\par
\begin{figure}[htbp]
\begin{center}
\includegraphics[scale=0.3]{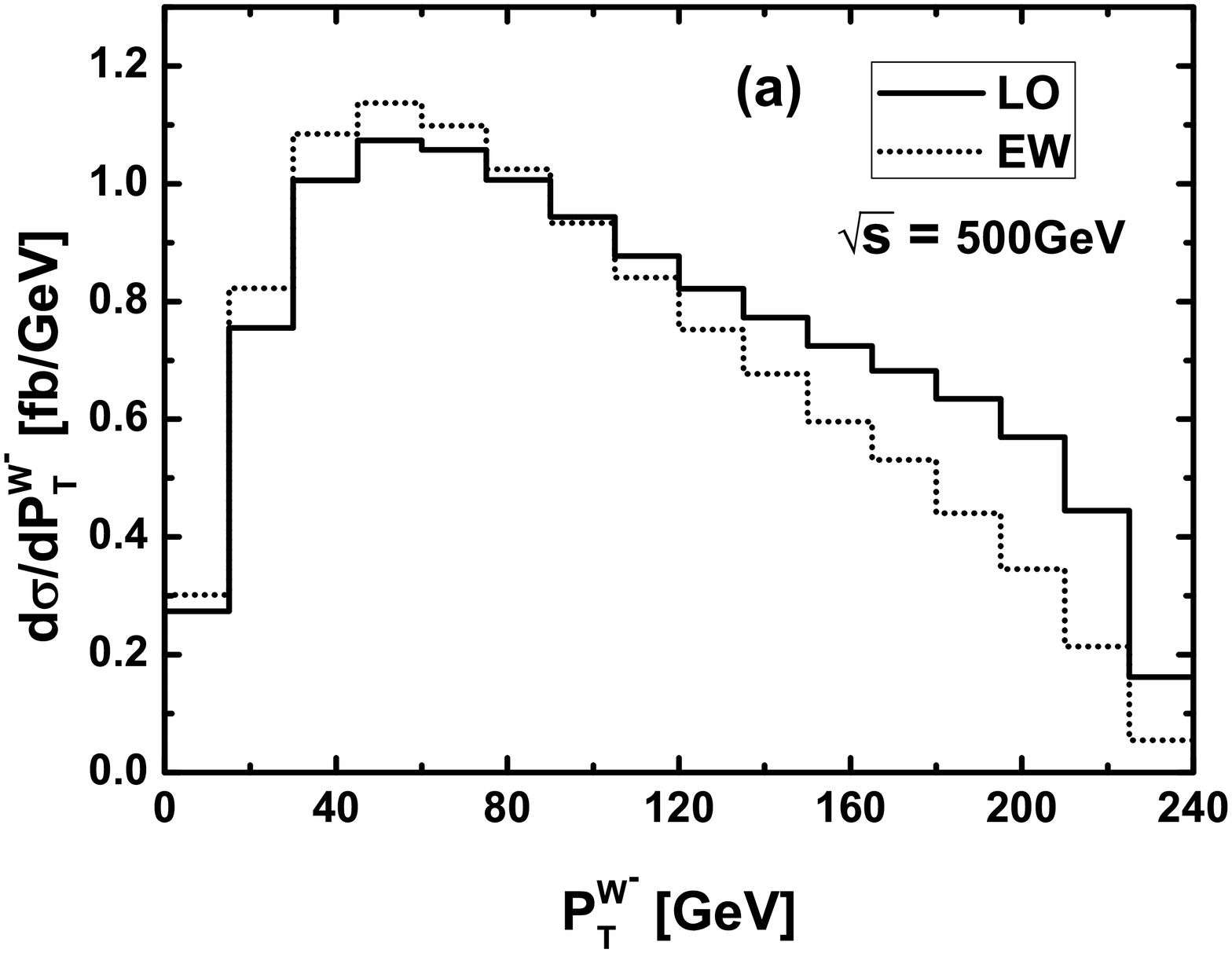}%
\includegraphics[scale=0.3]{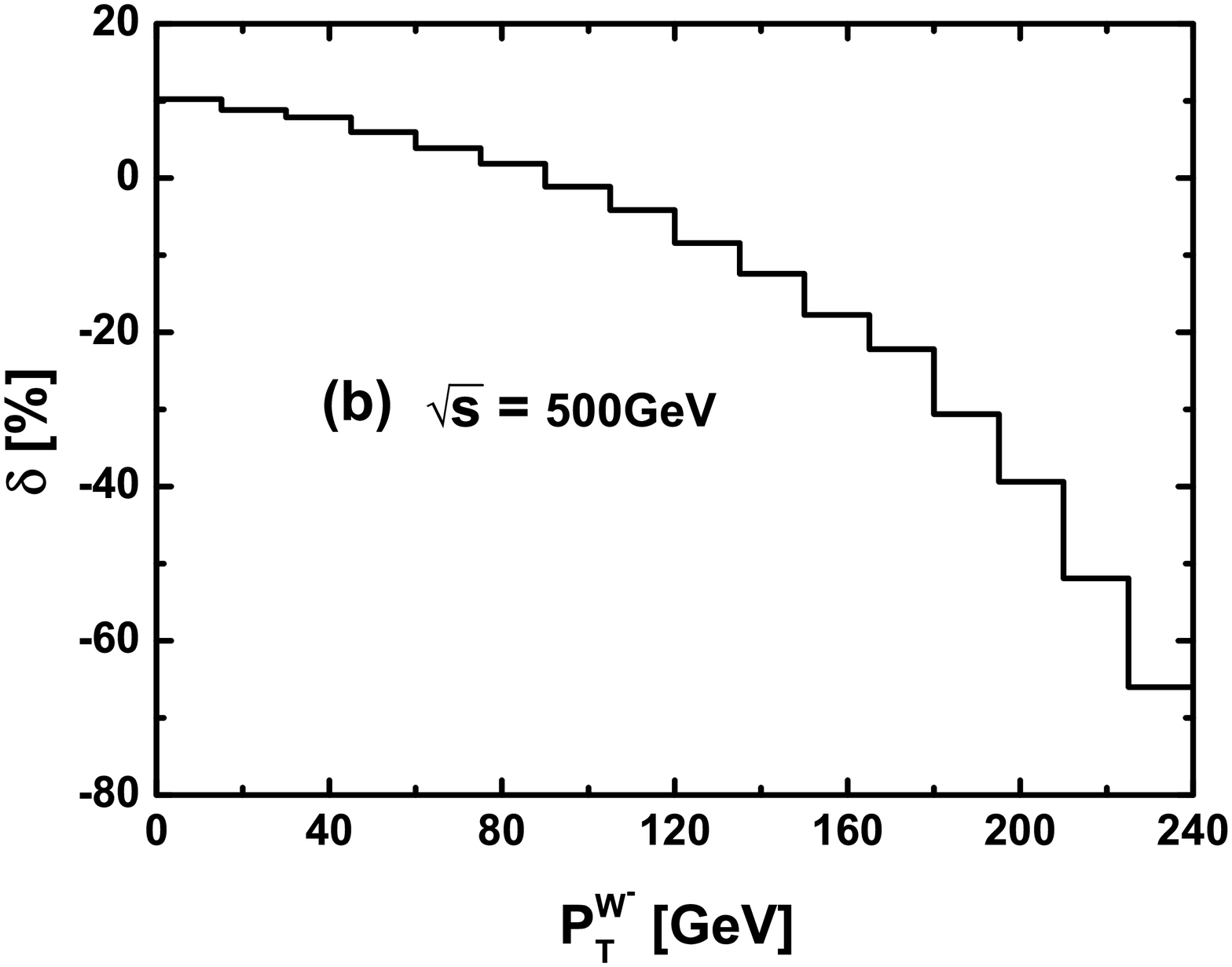}%
\caption{\label{ptw} (a) The LO and EW corrected transverse momentum distributions of the final $W^-$-boson for the \eewwr process in the SM. (b) The corresponding EW relative correction. }
\end{center}
\end{figure}
\begin{figure}[htbp]
\begin{center}
\includegraphics[scale=0.3]{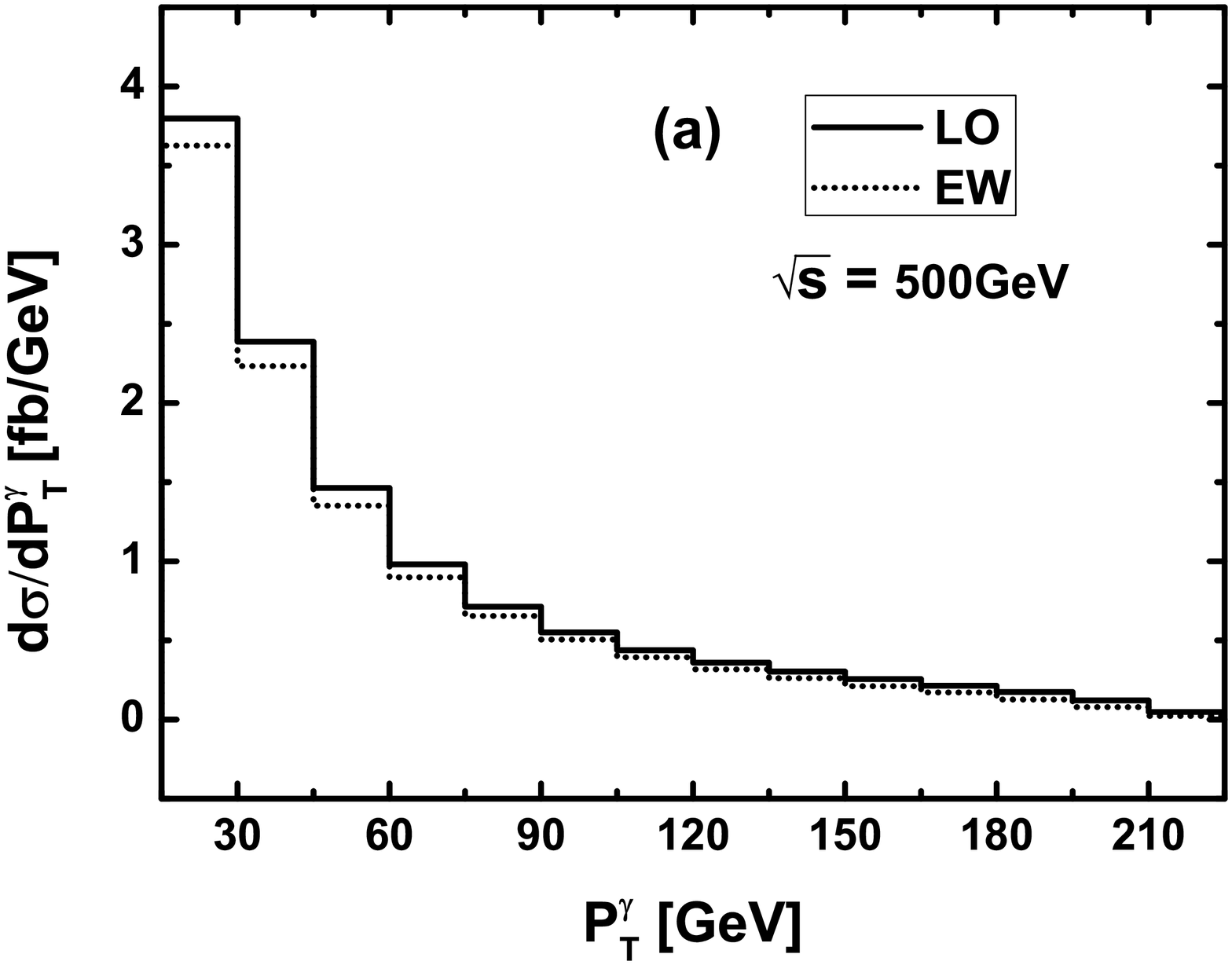}%
\includegraphics[scale=0.3]{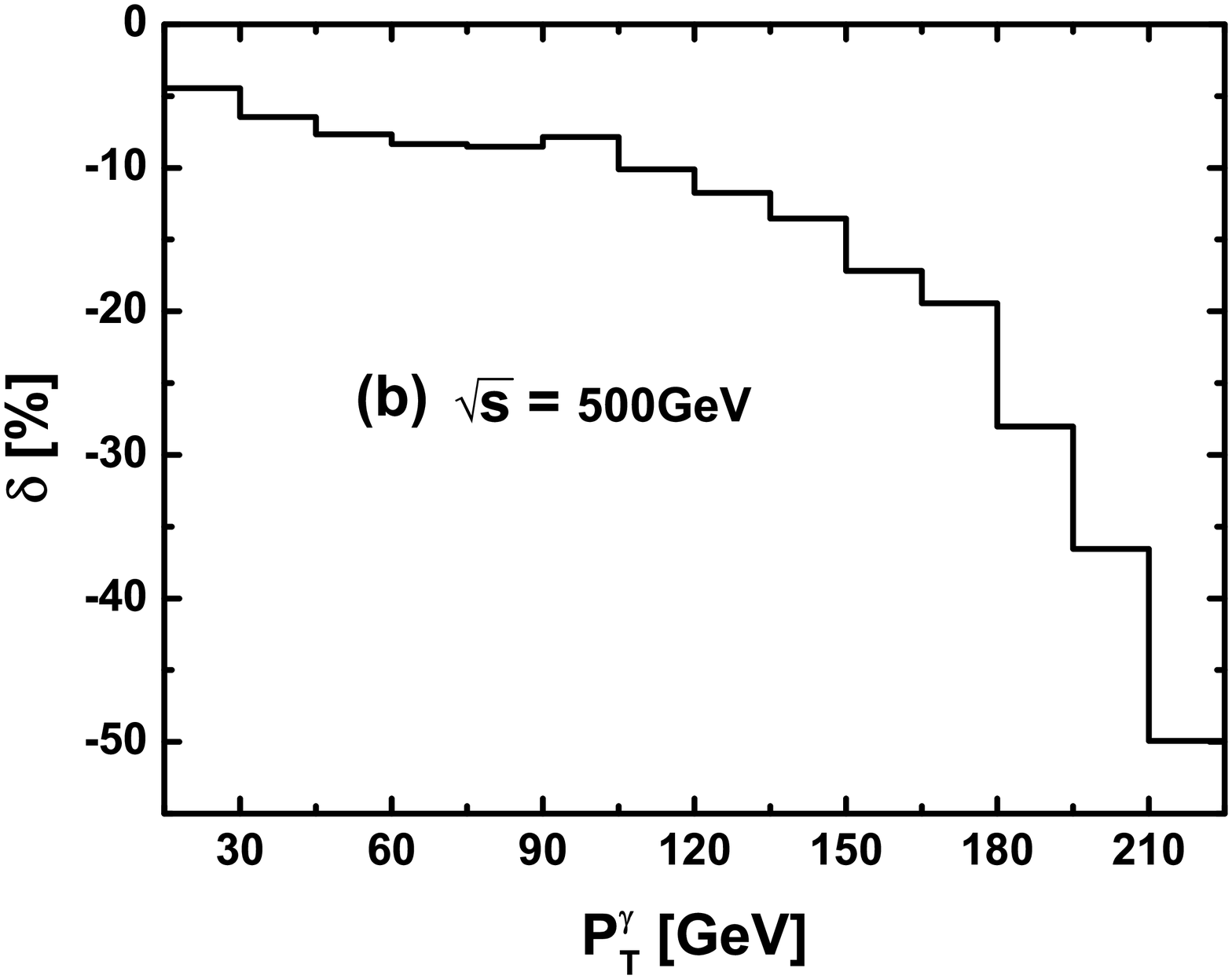}%
\caption{\label{ptr} (a) The LO and EW corrected transverse momentum distributions of the final leading photon for the \eewwr process in the SM. (b) The corresponding EW relative correction.}
\end{center}
\end{figure}

We plot the LO, EW corrected transverse momentum distributions of $W^-$-boson and the corresponding EW relative correction in Figs.\ref{ptw}(a) and (b), respectively. From the figures we can see that the EW correction enhances the LO differential cross section in the low $p_T^{W^-}$ region. With the increment of $p_T^{W^-}$, the relative EW correction goes down and changes from positive to negative at the position of $p_T^{W^-} \sim 90~{\rm GeV}$. The transverse momentum distributions of the final leading photon and the corresponding relative EW correction are shown in Figs.\ref{ptr}(a) and (b) separately. The kinematic cuts on the final photon have been declared before. As presented in Figs.\ref{ptr}(a, b), the LO and EW corrected distributions of $p_T^{\gamma}$ for the leading photon decrease violently with the increment of $p_T^{\gamma}$ and the corresponding EW correction suppresses the LO distribution in the whole plotted $p_T^{\gamma}$ region. Due to the Sudakov effect, the absolute EW relative corrections to the $p_T^{W^{-}}$ and $p_T^{\gamma}$ distributions become very large in high $p_T$ region.
%%%%%%%%%%%%%%fig-4%%%%%%%%%%%%%%%%%%%%%%%%%%%%%
\par
%%%%%%%%%%%%%%fig-4%%%%%%%%%%%%%%%%%%%%%%%%%%%%%
%%%%%%%%%%%%%%fig-5%%%%%%%%%%%%%%%%%%%%%%%%%%%%%
\par
%%%%%%%%%%%%%%fig-5%%%%%%%%%%%%%%%%%%%%%%%%%%%%%

\par
\begin{figure}[htbp]
\begin{center}
\includegraphics[scale=0.3]{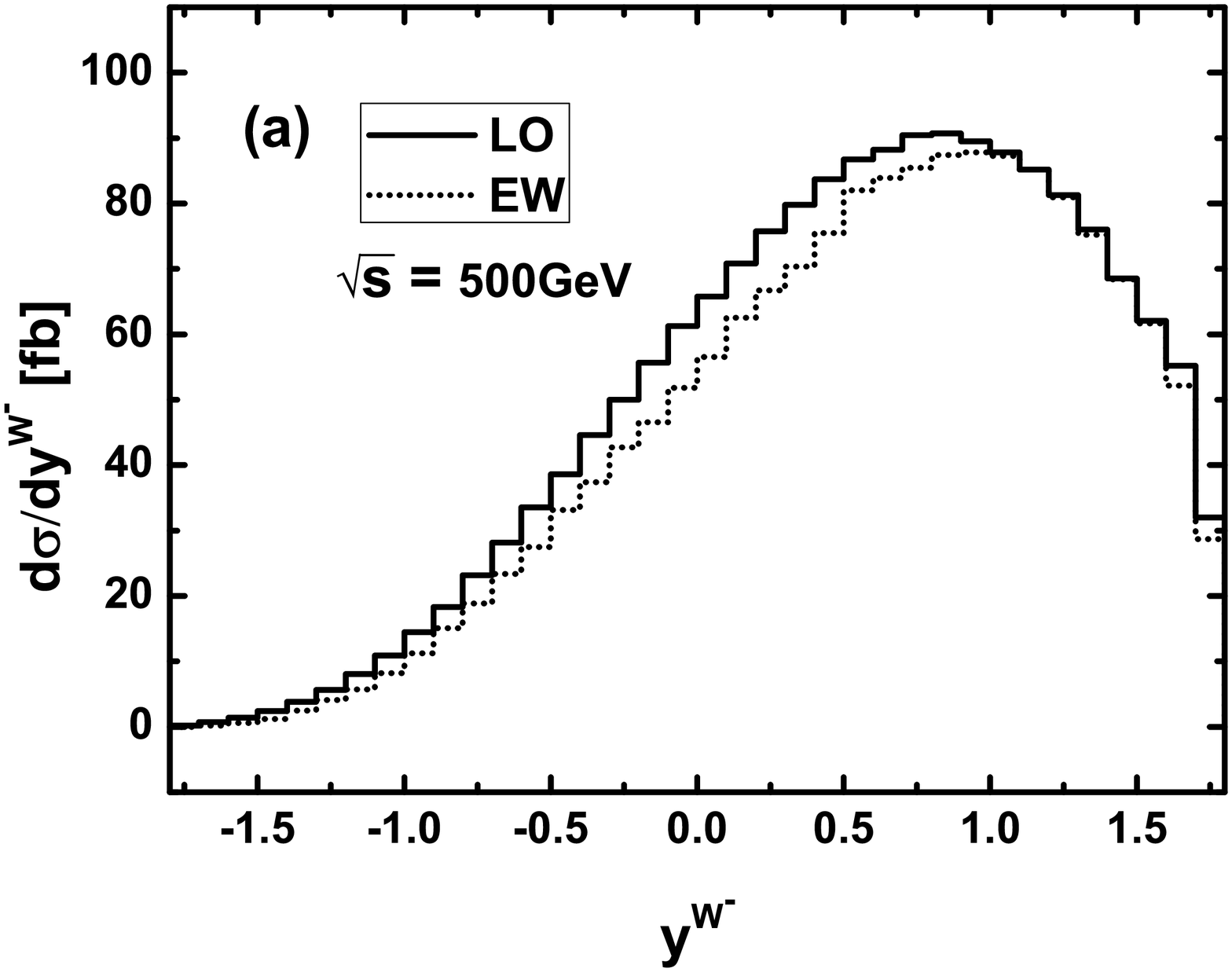}%
\includegraphics[scale=0.3]{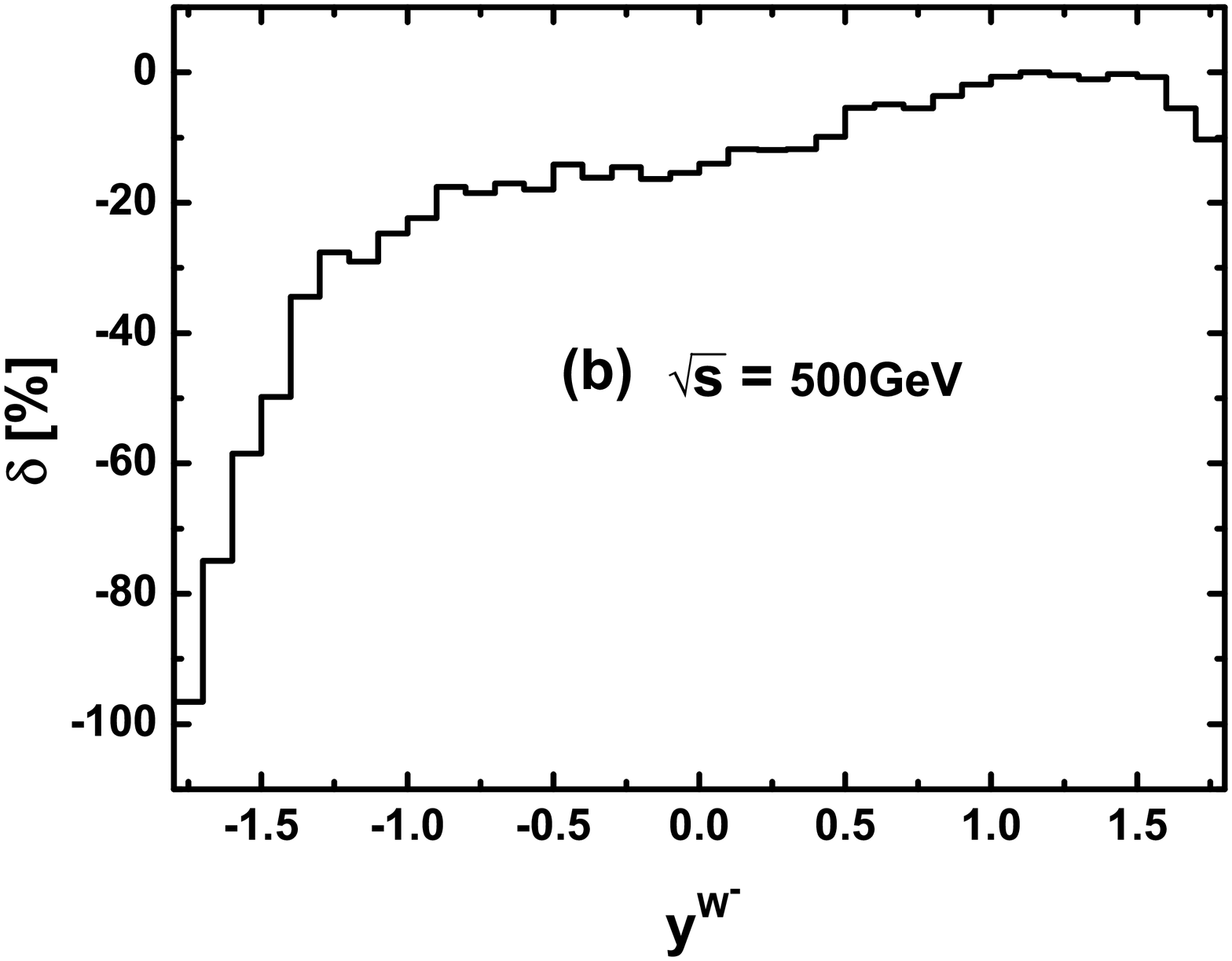}%
\caption{\label{yw} (a) The LO and EW corrected rapidity distributions of the final $W^-$-boson for the \eewwr process in the SM. (b) The corresponding EW relative correction.}
\end{center}
\end{figure}

The $z$-axis is defined in direction of the electron beam. The rapidity distributions of the final $W^-$-boson and the corresponding EW relative correction are presented in Figs.\ref{yw}(a) and (b), respectively. We find that the final $W^-$-boson tends to be forward with respect to the electron beam direction as shown in Fig.\ref{yw}(a). From the rapidity distributions of $W^-$-boson, we see that the LO and EW corrected differential cross sections reach their maxima in the vicinity of $y^{W^-} \sim 0.8$ and $y^{W^-} \sim 1$, respectively, and the EW correction suppresses the LO distribution in the whole plotted $y^{W^{-}}$ region. With the increment of $y^{W^-}$, the absolute EW relative correction decreases obviously.
%%%%%%%%%%%%%%fig-6%%%%%%%%%%%%%%%%%%%%%%%%%%%%%
\par
%%%%%%%%%%%%%%fig-6%%%%%%%%%%%%%%%%%%%%%%%%%%%%%

\par
\begin{figure}[htbp]
\begin{center}
\includegraphics[scale=0.3]{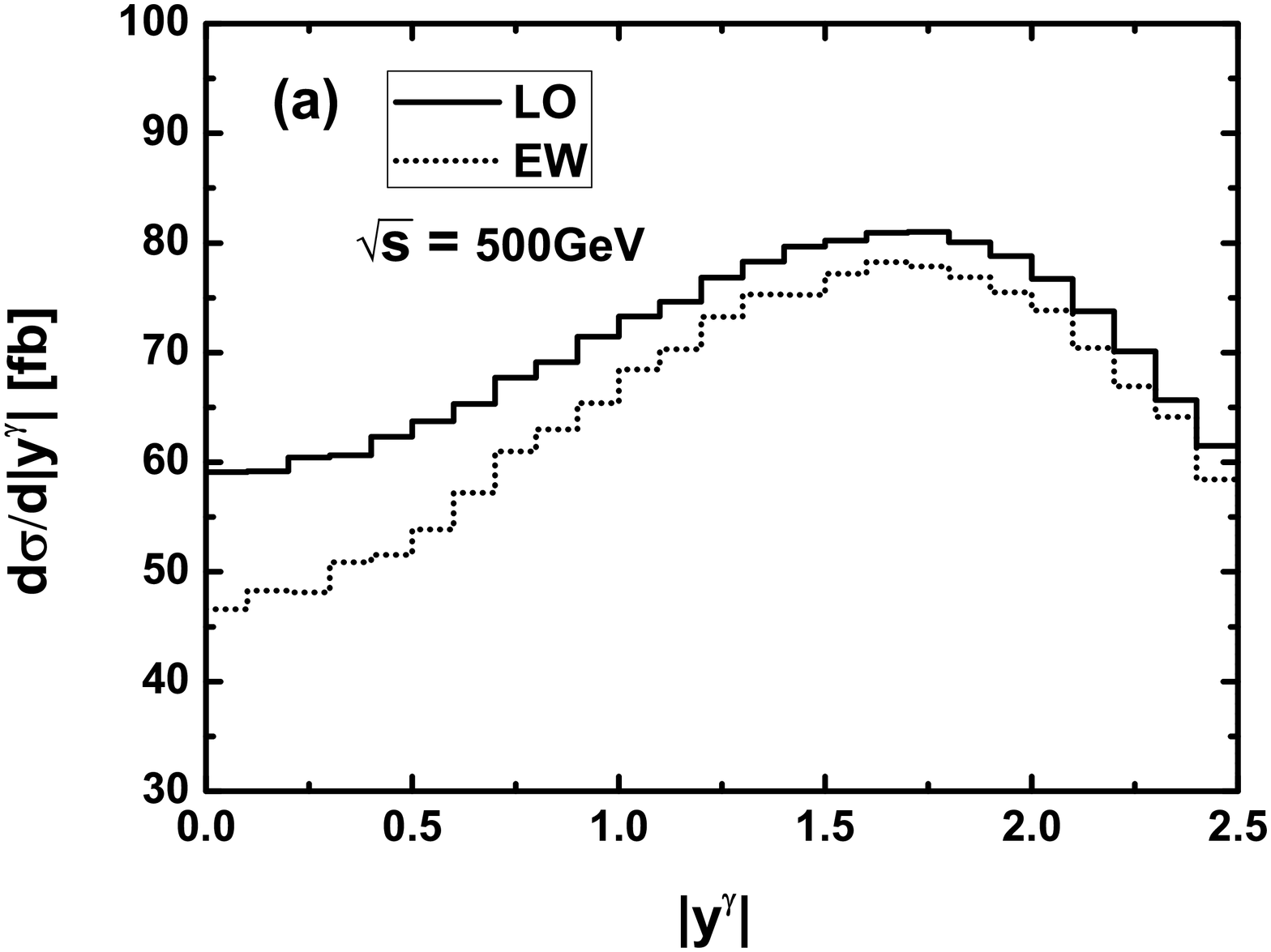}%
\includegraphics[scale=0.3]{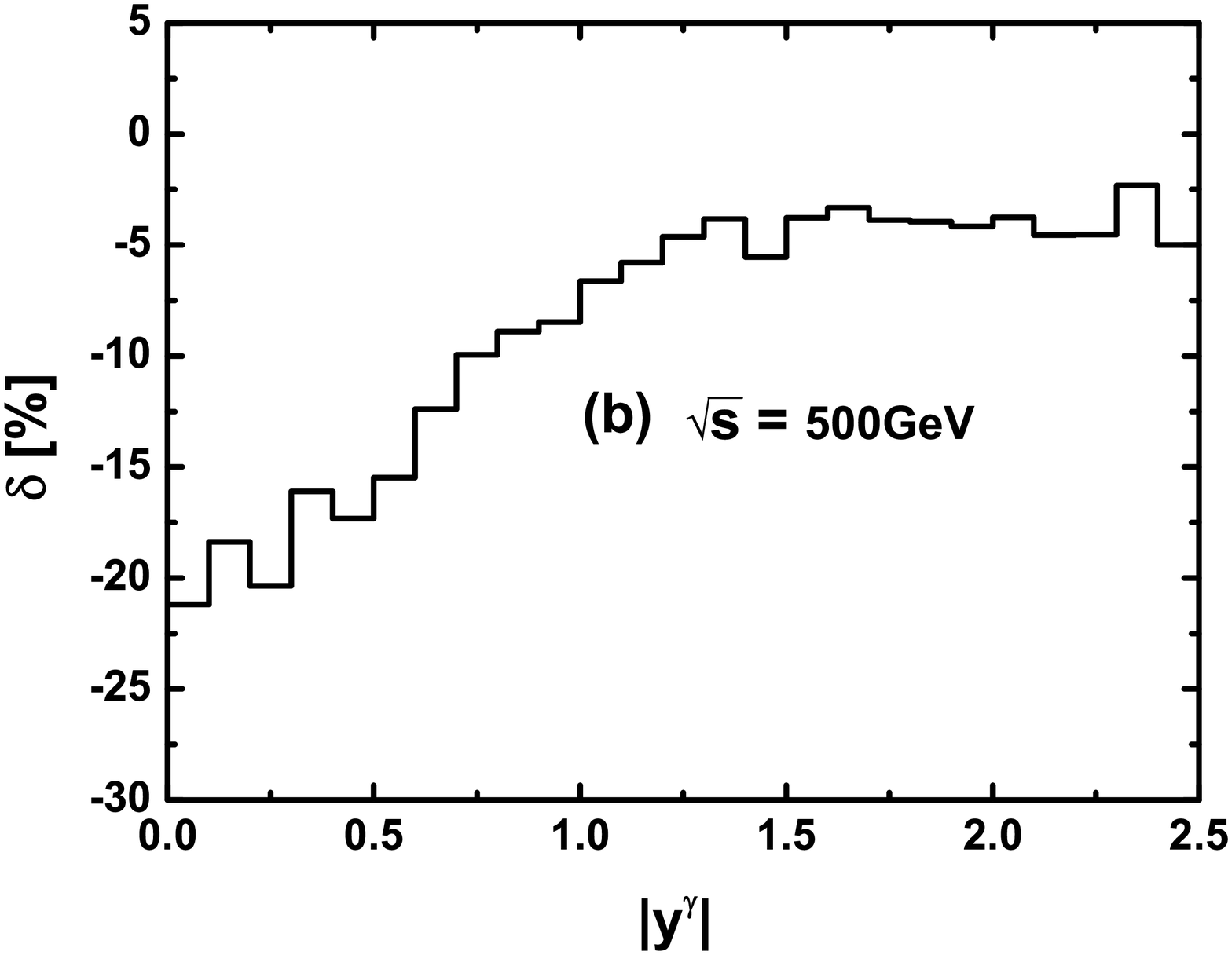}%
\caption{\label{yr} (a) The LO and EW corrected absolute rapidity distributions of the final leading photon for the \eewwr process in the SM. (b) The corresponding EW relative correction.}
\end{center}
\end{figure}

We plot the absolute rapidity distributions of the leading photon and the corresponding EW relative correction in Figs.\ref{yr}(a) and (b), separately. From the figures we can see that the LO and EW corrected absolute rapidity distributions of the leading photon peak at the position of $\vert y^{\gamma} \vert \sim 1.8$ and the EW correction suppresses the LO distribution in the whole plotted $|y^{\gamma}|$ region. We see also that the absolute EW relative correction to rapidity distribution of the leading photon reaches its maximum in the central rapidity region, i.e., $y^{\gamma} = 0$.
%%%%%%%%%%%%%%fig-7%%%%%%%%%%%%%%%%%%%%%%%%%%%%%
\par
%%%%%%%%%%%%%%fig-7%%%%%%%%%%%%%%%%%%%%%%%%%%%%%

\par
\begin{figure}[htbp]
\begin{center}
\includegraphics[scale=0.3]{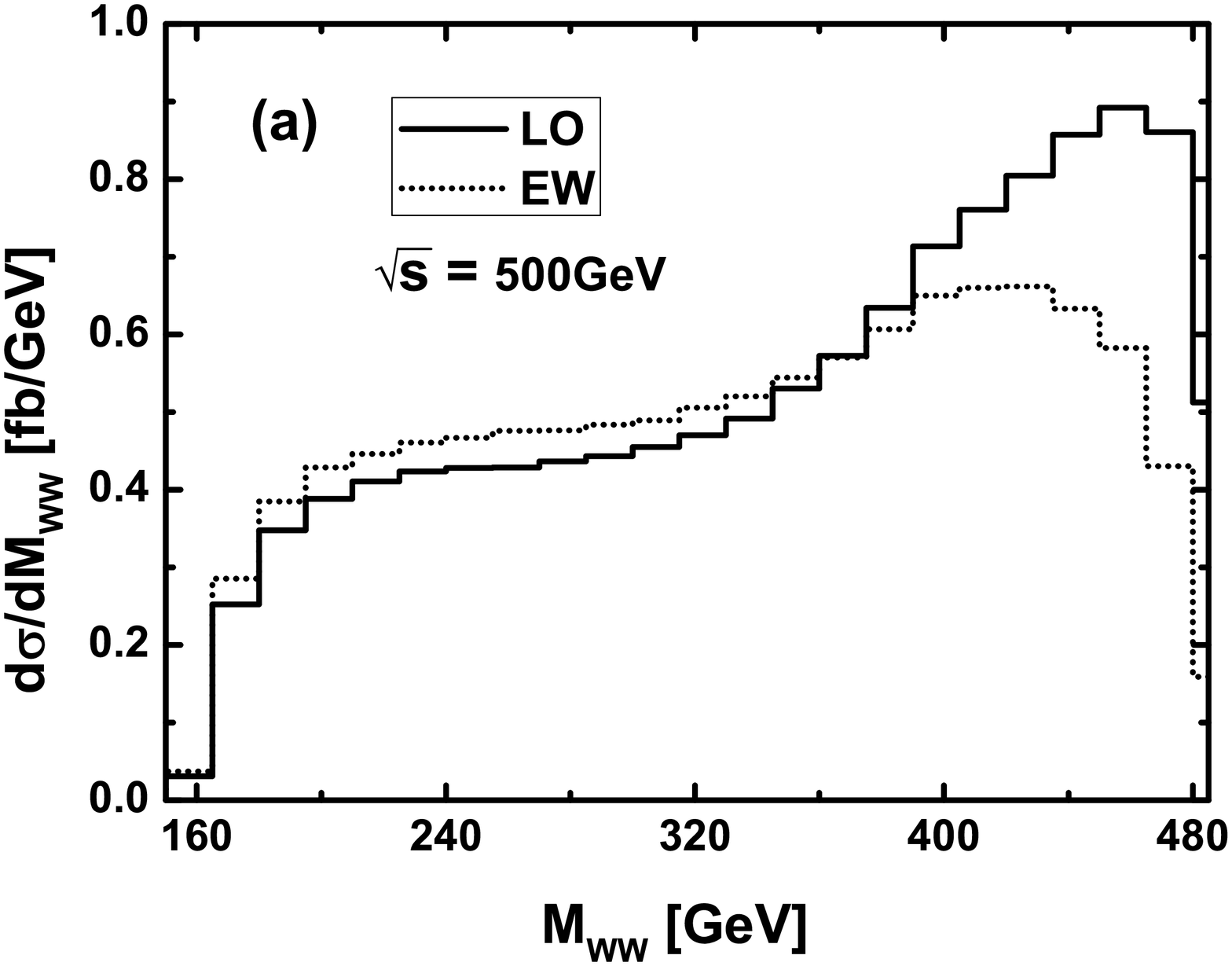}%
\includegraphics[scale=0.3]{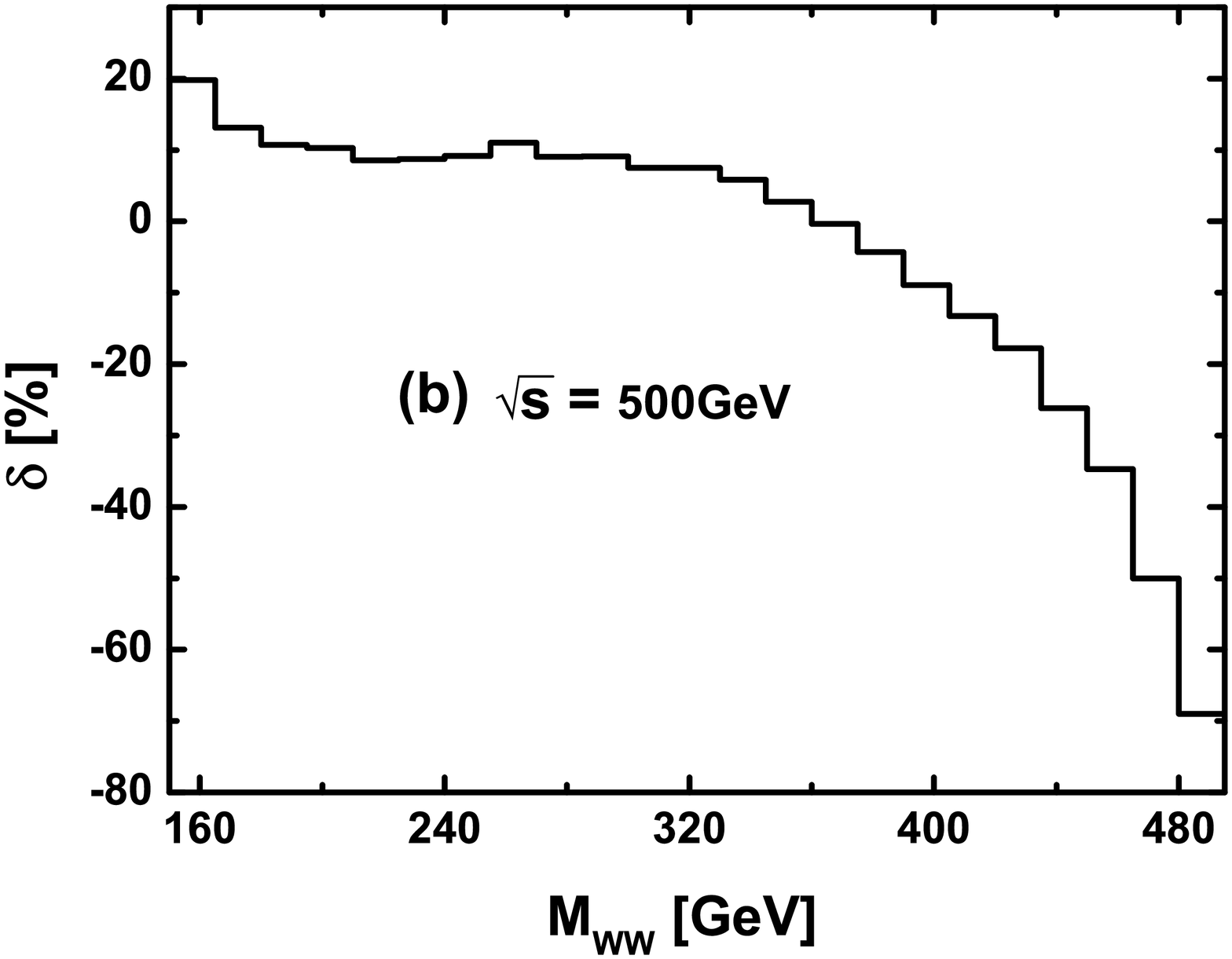}%
\caption{\label{mww} (a) The LO and EW corrected distribution of the invariant mass $M_{WW}$ for the \eewwr process in the SM. (b) The corresponding EW relative correction.}
\end{center}
\end{figure}

In Figs.\ref{mww}(a) and (b), we depict the LO and EW corrected distributions of $W$-pair invariant mass $M_{WW}$ and the corresponding EW relative correction, separately. The differential cross sections of $M_{WW}$ are drawn in the range of $M_{WW} \in [2M_W,485~{\rm GeV}]$, where the upper limit on $M_{WW}$ is determined by the colliding energy and the transverse momentum cut on the final photon. Fig.\ref{mww}(a) shows that in the range of $400~{\rm GeV} > M_{WW} > 180~{\rm GeV}$ the LO and EW corrected cross sections go up obviously. The EW correction enhances the LO distribution in small $M_{WW}$ region, and the EW relative correction decreases from positive to negative at the position of $M_{WW} \sim 360~ {\rm GeV}$ with the increment of $M_{WW}$. The obvious EW correction in large $M_{WW}$ region can be ascribed to the Sudakov logarithms originating from the virtual contribution.
%%%%%%%%%%%%%%fig-8%%%%%%%%%%%%%%%%%%%%%%%%%%%%%
\par
%%%%%%%%%%%%%%fig-8%%%%%%%%%%%%%%%%%%%%%%%%%%%%%

\par
Following the definition in Ref.\cite{FB-A}, we have the expression for the forward-backward asymmetry of $W^-$-boson as
\begin{eqnarray}
A_{W^-}=\frac{\sigma(y^{W^-}>0)~-~\sigma(y^{W^-}<0)}{\sigma(y^{W^-}>0)~+~\sigma(y^{W^-}<0)}~.
\end{eqnarray}
We calculate the LO and EW corrected forward-backward asymmetries and obtain $A_{W^-}^{LO}=~54.72\%$, $A_{W^-}^{EW}=~59.44\%$, separately. These numerical results show that both the $A_{W^-}^{LO}$ and $A_{W^-}^{EW}$ are significant and most of the $W^-$-bosons are produced in the forward hemisphere. That feature is also shown in Fig.\ref{yw}(a).

\par
\begin{figure}[htbp]
\begin{center}
\includegraphics[scale=0.3]{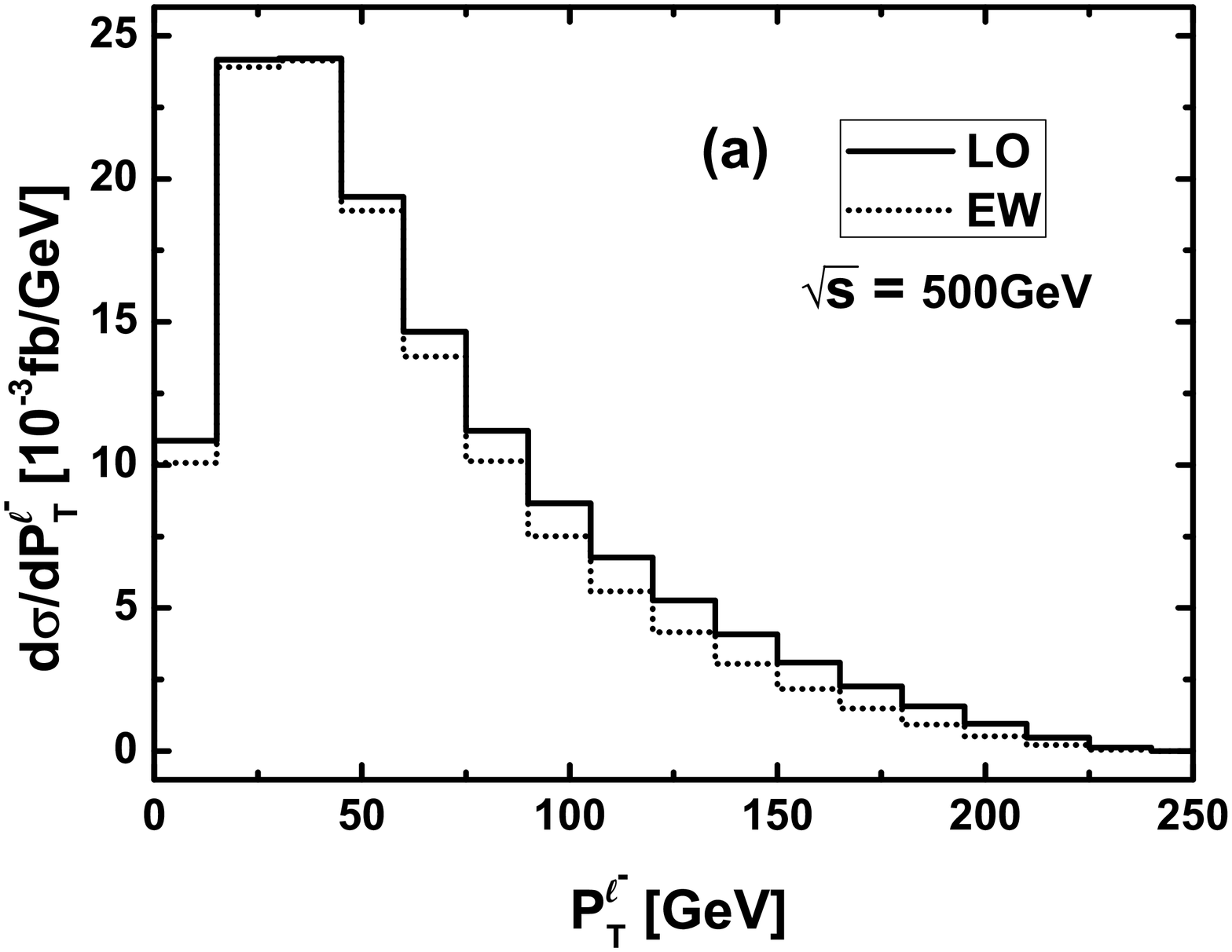}%
\includegraphics[scale=0.3]{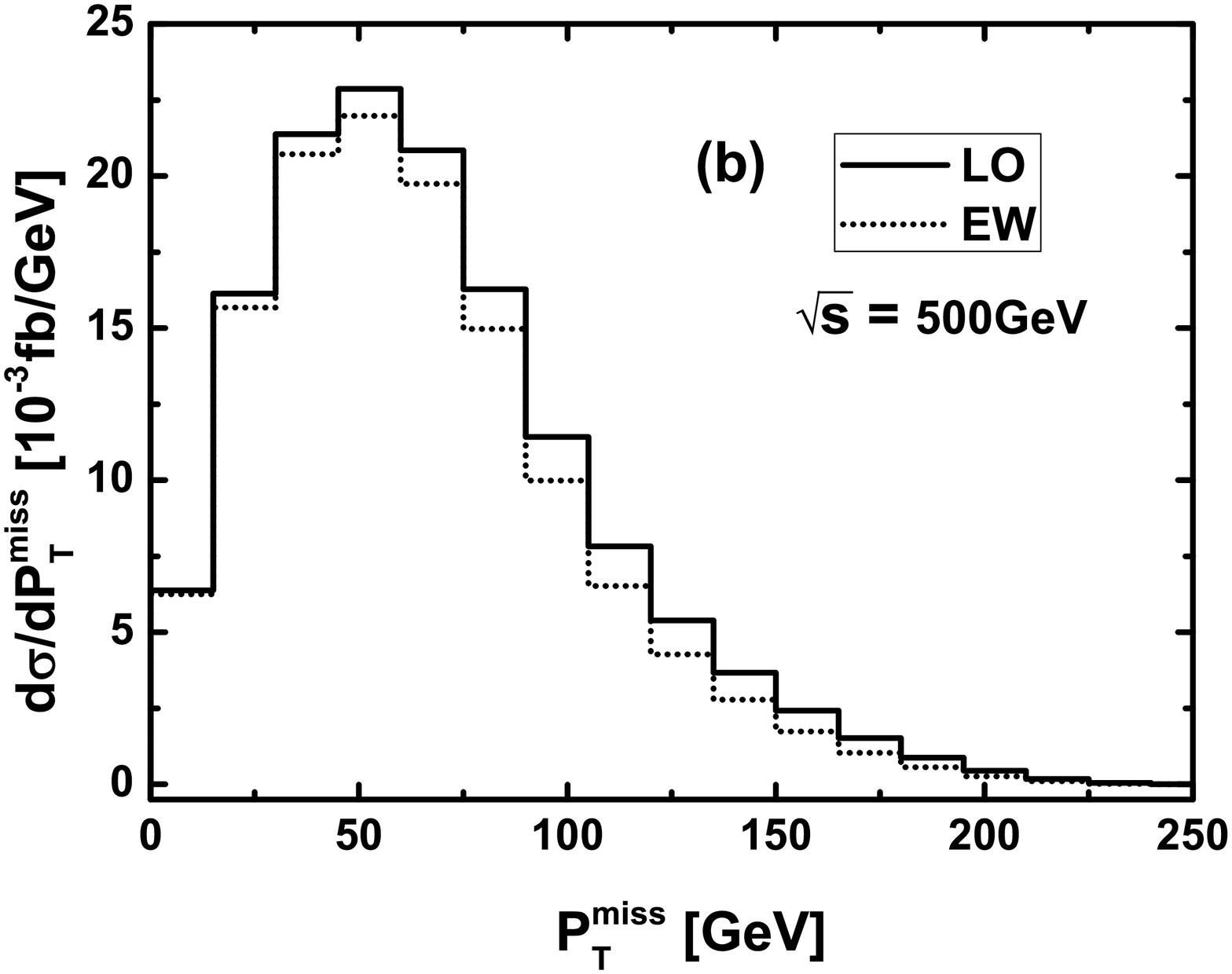}%
\caption{\label{decay1} The LO and EW corrected distributions of (a) $p_T^{l^{-}}$ and (b) $p_T^{miss}$ for the $e^+ e^- \to W^+ W^- \gamma \to l_1^+ l_2^- \nu_{l_1} \bar{\nu}_{l_2} \gamma$ ($l_1, l_2 = e, \mu$) process in the SM.}
\end{center}
\end{figure}
\begin{figure}[htbp]
\begin{center}
\includegraphics[scale=0.3]{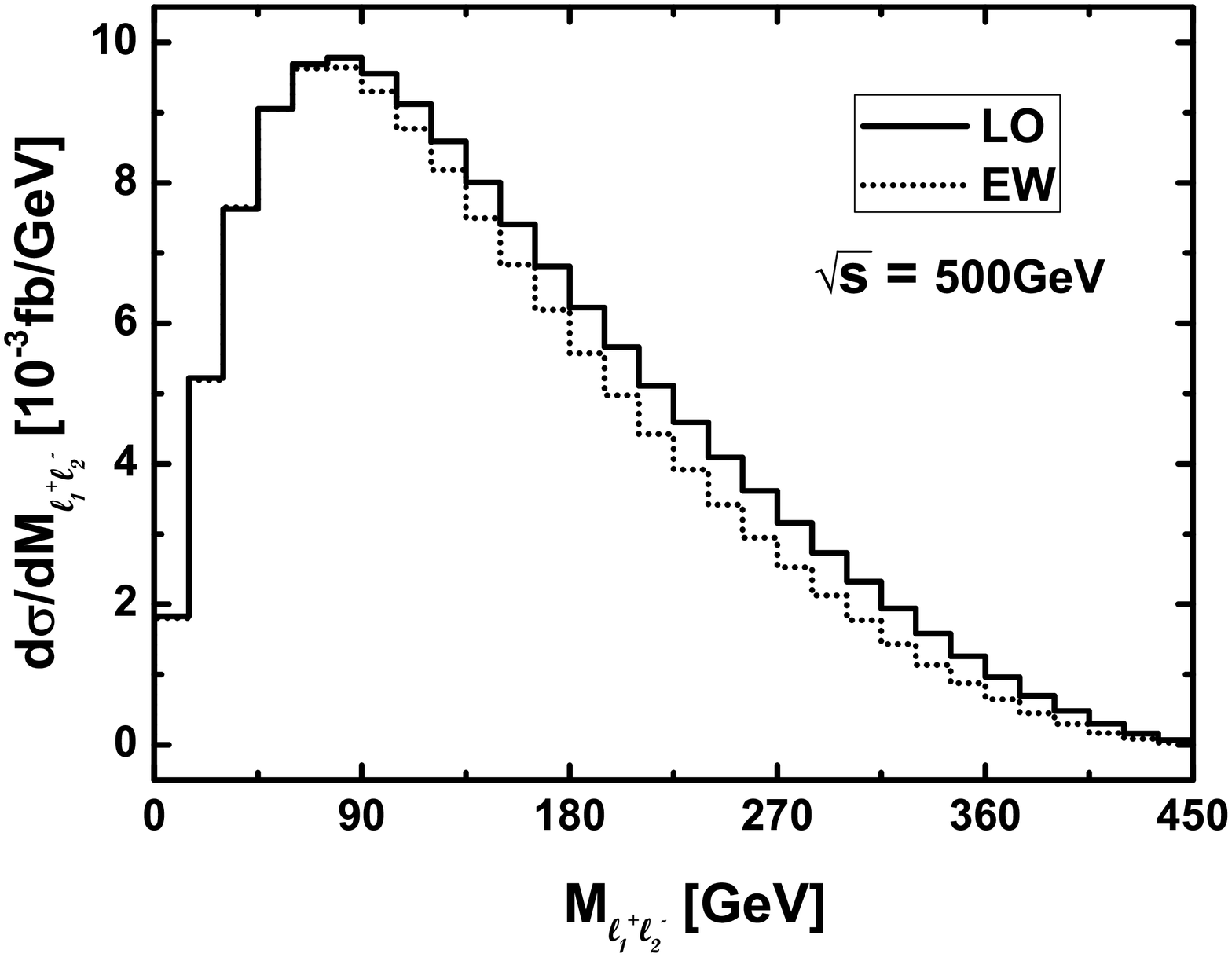}%
\caption{\label{decay2} The LO and EW corrected distributions of the lepton pair invariant mass $M_{l_1^+l_2^-}$ for the \decayeewwr ($l_1,l_2=e,\mu$) process in the SM.}
\end{center}
\end{figure}

Now we consider the leptonic decays of the final $W$-boson pair by neglecting the mass difference between the electron and muon and adopting the narrow width approximation (NWA). We take the relevant branch ratio as $Br(W\to l\nu)=10.80\%~(l=e,~\mu)$ \cite{pdg}, and depict the LO and EW corrected distributions of the final negative charged lepton transverse momentum and missing transverse momentum in Figs.\ref{decay1}(a) and (b), respectively. From the figures we can see that the distributions reach their maxima at the position of $p_T^{l^-} \sim 30~{\rm GeV}$ and $p_T^{miss} \sim 50~{\rm GeV}$, separately, and the EW correction suppresses the LO distributions in the whole plotted $p_T$ region. Furthermore, we present the distributions of the lepton pair invariant mass $M_{l_1^+l_2^-}$ ($l_1,l_2=e,\mu$) in Fig.\ref{decay2}. The figure shows that both the LO and EW corrected lepton pair invariant mass distributions reach their maxima in the vicinity of $M_{l_1^+l_2^-} \sim 80~{\rm GeV}$, and the EW correction suppresses the LO distribution in the whole plotted lepton pair invariant mass region. Particularly, we see that in the region of $M_{l_1^+l_2^-} > 90~{\rm GeV}$, the EW correction suppresses the LO cross section significantly.
%%%%%%%%%%%%%%fig-9%%%%%%%%%%%%%%%%%%%%%%%%%%%%%
\par
%%%%%%%%%%%%%%fig-9%%%%%%%%%%%%%%%%%%%%%%%%%%%%%
%%%%%%%%%%%%%%fig-10%%%%%%%%%%%%%%%%%%%%%%%%%%%%%
\par
%%%%%%%%%%%%%%fig-9%%%%%%%%%%%%%%%%%%%%%%%%%%%%%

\par
\begin{figure}[htbp]
\begin{center}
\includegraphics[scale=0.3]{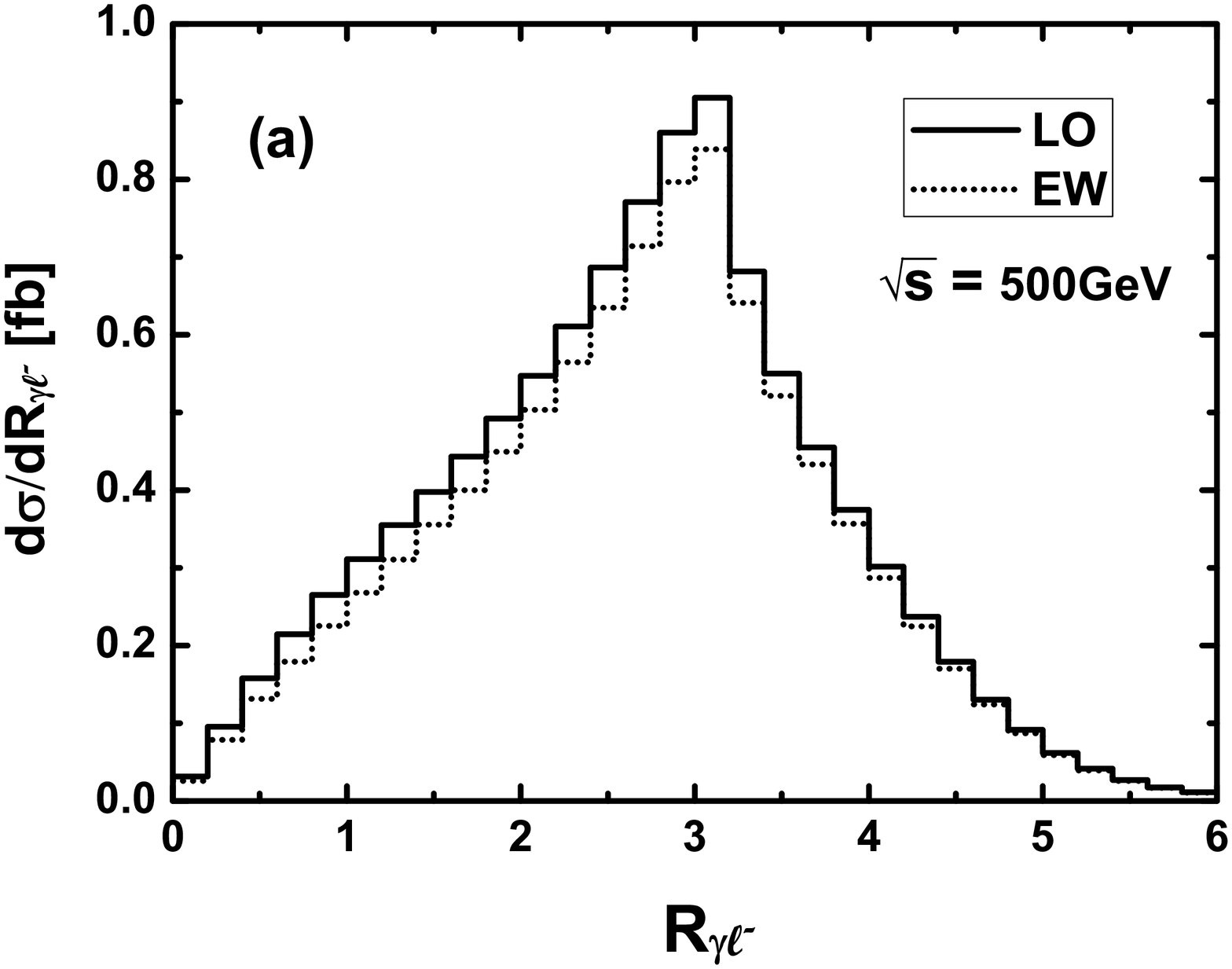}%
\includegraphics[scale=0.3]{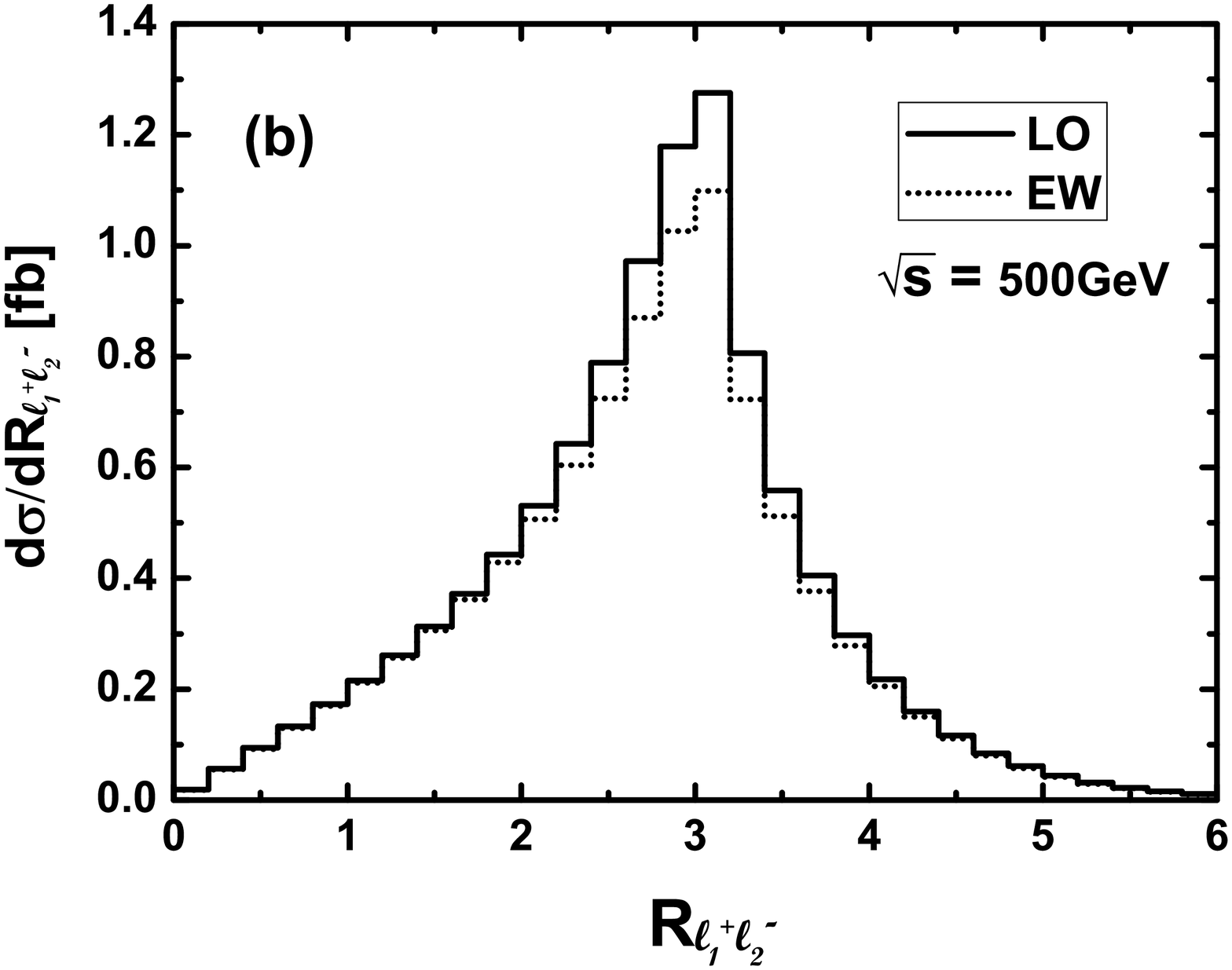}%
\caption{\label{decay-R} The distributions for the \decayeewwr ($l_1,l_2=e,\mu$) process in the SM with the colliding energy $\sqrt s~=~500~{\rm GeV}$. (a) $R_{\gamma l^-}$ distributions. (b) $R_{l_1^+ l_2^-}$ distributions.}
\end{center}
\end{figure}

The LO and EW corrected distributions of the photon-lepton and lepton-lepton separations in the rapidity-azimuthal-angle plane, $R_{\gamma l^-}$ and $R_{l_1^+ l_2^-}$, are depicted in Figs.\ref{decay-R}(a) and (b), respectively. The figures show that the final photon and leptons are well separated and the $R_{\gamma l^-}$, $R_{l_1^+l_2^-}$ distributions reach their maxima at the position of $R \sim 3$. We also observe an obvious dependence of the EW correction on the value of $R$. The $R_{\gamma l^-}$ distributions demonstrate that the EW correction is more obvious in the relatively small $R_{\gamma l^-}$ region, while the $R_{l_1^+ l_2^-}$ distributions show that the EW correction concentrates in the vicinity of $R_{l_1^+ l_2^-} \sim 3$.
%%%%%%%%%%%%%%fig-11%%%%%%%%%%%%%%%%%%%%%%%%%%%%%
\par
%%%%%%%%%%%%%%fig-11%%%%%%%%%%%%%%%%%%%%%%%%%%%%%

\vskip 5mm
\section{Summary}
\par
In this paper, we present the full NLO EW corrections and the high order initial state radiation contributions in the leading-logarithmic approximation to the $W^+W^-\gamma$ production in $e^+e^-$ collision mode at the ILC. The \eewwr process involves the $W^+W^-\gamma \gamma$ and $W^+W^-Z\gamma~$ QGCs at the tree level, thus it is very important not only in exploring the non-Abelian structures of the SM, but also in identifying the electroweak symmetry breaking mechanism. Our results show that the EW correction suppresses the LO cross section significantly and the EW relative correction varies in the region of $[-26.62\%, -8.05\%]$ when $\sqrt s$ goes up from $180~{\rm GeV}$ to $1~{\rm TeV}$. We find that near the threshold the ISR effect beyond ${\cal O}(\alpha)$ is important, while at the high colliding energy region it is small and negligible. We also plot the LO and EW corrected differential cross sections of $p_T^{W^-,\gamma}$, $y^{W^-,\gamma}$ and $M_{WW}$. From the various kinematic variable distributions, we find a strong phase dependence of the EW correction. Finally, we investigate the leptonic decays of the final $W$-boson pair by adopting the NWA. The results show that by adopting our event selection criteria the final photon and leptons can be well separated and the $R_{\gamma l^-,l_1^+l_2^-}$ ($l,l_1,l_2=e,\mu$) distributions reach their maxima at the position of $R \sim 3$.

\par
\section{Acknowledgments}
This work was supported in part by the National Natural Science Foundation of China (Grant No.11275190, No.11375008, No.11375171).

\end{document}